\documentclass[manuscript,screen]{acmart}

\AtBeginDocument{%
  \providecommand\BibTeX{{%
    \normalfont B\kern-0.5em{\scshape i\kern-0.25em b}\kern-0.8em\TeX}}}

\setcopyright{acmlicensed}
\usepackage{graphicx} % Required for inserting images
\usepackage{amsmath,verbatim,mathrsfs,latexsym,paralist}
\usepackage{indentfirst}
\usepackage{color}
\usepackage{amsthm}
\usepackage{xspace}
\newtheorem{Thm}{Theorem}
\newtheorem{Def}{Definition}
\usepackage{algorithm}

\usepackage{algorithm}
\usepackage{algpseudocode}
\usepackage{mathtools}
\usepackage[colorinlistoftodos,prependcaption,textsize=tiny]{todonotes}
\usepackage{calrsfs}
\usepackage{dutchcal}

\begin{document}

%%
%% The "title" command has an optional parameter,
%% allowing the author to define a "short title" to be used in page headers.
\title{Post-hoc Evaluation of Nodes Influence in Information Cascades: The Case of Coordinated Accounts}

\author{Niccolò Di Marco}
\email{niccolo.dimarco@uniroma1.it}
\orcid{0000-0003-4335-7328}
\affiliation{%
  \institution{Sapienza University of Rome}
  \streetaddress{Viale Regina Elena 295}
  \city{Roma}
  \country{Italy}
}

\author{Sara Brunetti}
\email{sara.brunetti@unisi.it}
\orcid{0000-0001-8607-6900}
\affiliation{%
  \institution{University of Siena}
  % \streetaddress{Viale Regina Elena 295}
  \city{Siena}
  \country{Italy}
}

\author{Matteo Cinelli}
\email{matteo.cinelli@uniroma1.it}
\orcid{0000-0003-3899-4592}
\affiliation{%
  \institution{Sapienza University of Rome}
  \streetaddress{Viale Regina Elena 295}
  \city{Roma}
  \country{Italy}
}

\author{Walter Quattrociocchi}
\email{walter.quattrociocchi@uniroma1.it}
\orcid{0000-0002-4374-9324}
\affiliation{%
  \institution{Sapienza University of Rome}
  \streetaddress{Viale Regina Elena 295}
  \city{Roma}
  \country{Italy}
}

%%
%% By default, the full list of authors will be used in the page
%% headers. Often, this list is too long, and will overlap
%% other information printed in the page headers. This command allows
%% the author to define a more concise list
%% of authors' names for this purpose.
% \renewcommand{\shortauthors}{Trovato and Tobin, et al.}

%%
%% The abstract is a short summary of the work to be presented in the
%% article.
\begin{abstract}
In the last years, social media has gained an unprecedented amount of attention, playing a pivotal role in shaping the contemporary landscape of communication and connection. However, Coordinated Inhautentic Behaviour (CIB), defined as orchestrated efforts by entities to deceive or mislead users about their identity and intentions, has emerged as a tactic to exploit the online discourse. 
In this study, we quantify the efficacy of CIB tactics by defining a general framework for evaluating the influence of a subset of nodes in a directed tree.
We design two algorithms that provide optimal and greedy post-hoc placement strategies that lead to maximising the configuration influence. 
We then consider cascades from information spreading on Twitter to compare the observed behaviour with our algorithms.
The results show that, according to our model, coordinated accounts are quite inefficient in terms of their network influence, thus suggesting that they may play a less pivotal role than expected. 
Moreover, the causes of these poor results may be found in two separate aspects: a bad placement strategy and a scarcity of resources.
\end{abstract}

%%
%% The code below is generated by the tool at http://dl.acm.org/ccs.cfm.
%% Please copy and paste the code instead of the example below.
%%


%%
%% Keywords. The author(s) should pick words that accurately describe
%% the work being presented. Separate the keywords with commas.

% \received{20 February 2007}
% \received[revised]{12 March 2009}
% \received[accepted]{5 June 2009}

\maketitle
\section{Introduction}

Social media has transformed how we connect and share information, thus reshaping some of the mechanisms that we use for engaging with the world.
Platforms that have been initially designed for entertainment, are increasingly becoming the principal environment in which opinions and views take form, determining a new tendency in information consumption \cite{bakshy2012role, flaxman2016filter, schmidt2017anatomy}. 

However, this radical change came at the cost of a number of downsides such as online harassment, toxicity and hateful speech \cite{ZINOVYEVA2020113362,trujillo2022make}, polarisation \cite{garrett2009echo,bail2018exposure,tucker2018social,cinelli2021echo,falkenberg2022growing} and the spreading of misleading information \cite{bakshy2012role, del2016spreading, lazer2018science, bovet2019influence,santos2021link, hristakieva2022spread,juul2021comparing,pierri2023propaganda}. 
In this landscape, special attention has been posed to investigate the phenomenon of {\it Coordinated Inauthentic Behaviour} (CIB) which, according to Meta's definition, is "the use of multiple Facebook or Instagram assets, working in concert to engage in inauthentic behaviour, where the use of fake accounts is central to the operation". More in detail, according to the platform's Community Standards (\url{https://transparency.fb.com/en-gb/policies/community-standards/inauthentic-behavior/}), the concept of {\it inauthentic behaviour} refers to people who "misrepresent themselves on Facebook, use fake accounts, artificially boost the popularity of content or engage in behaviours designed to enable other violations". Related to CIB researchers investigated the phenomenon of {\it Coordinated Behavior} (CB), which can be defined as an unexpected, suspicious, or exceptional similarity among users of a group~\cite{nizzoli2021coordinated}. 

Recent studies highlight that accounts displaying CIB, CB and social bots (i.e. software agents that communicate autonomously on social media and other platforms, having different uses and purposes \cite{Lebeuf2019,Seering2018,Zheng2019,Klopfenstein2017}) may have played a role in relevant events~\cite{ruths2019misinformation} such as political elections~\cite{grinberg2019fake, cinelli2022coordinated, keller2020political}, and in disseminating false information~\cite{shao2018spread,mendoza2020bots, cresci2020decade, sharma2021identifying, Pacheco2020}. It follows that, beyond detection challenges~\cite{nizzoli2021coordinated, pacheco2021uncovering, luceri2023unmasking, nwala2023language}, one key reason to study such accounts relies on their potential to manipulate public opinion by swaying the narrative and influence the perceptions and views of a large audience \cite{schoch2022coordination}.

In this work, we measure the influence of such potentially malicious accounts as the numerosity of the audience they interact with. In particular, we introduce a general framework that allow us to compare their influence with two different theoretical (post-hoc) models for coordinated accounts placement in information cascades: 

\begin{enumerate}
    \item first we consider an optimal model, in which we identify the number and disposition of coordinated accounts to ensure the maximal influence over the tree, without imposing any constraints on the number of coordinated accounts to be used;
    \item second, we focus on a greedy strategy in which we have a fixed number of coordinated accounts to place to maximise the influence (i.e. limited capabilities case).
\end{enumerate}

Notably, the algorithms provided can be extended to any other influence problem modelled using binary labels on trees.

Using simulations on synthetic data, we show that, on average, it is possible to exert maximum influence using a limited number of coordinated accounts whose value highly depends on the height of the considered tree. Furthermore, the labelling that corresponds to the optimal placement results to be rare compared to the whole set of possible arrangements of node labels. 
As a case study, we compare the influence obtained by our models in a large dataset made of $\sim50K$ Twitter cascades, built starting from $~1.4M$ of tweets about the 2019 UK political elections \cite{cinelli2022coordinated}.

Our results show that, according to our modelling, coordinated accounts exert a much lower influence than the one obtained using both the unconstrained algorithm and the greedy strategy. The reasons for this behaviour have to be searched in two distinct factors: a limited number of available resources (i.e. coordinated accounts) and a bad placement strategy. 
We conclude by noticing that, in general, the observed placement closely resembles a random placement rather than a specific strategy.

Despite the limitations of our post-hoc models, our results suggest that CB may exert a limited influence, contrary to what is expected.

\section{Definitions}

We consider a directed tree $T = (V,E)$ such that $|V| = n$. In particular, each node represents a user, while a directed edge indicates the direction of the flow of information among them. To distinguish between coordinated and non-coordinated users, we assign a binary label $\mathcal{l}_v$ to each node, such that $v$ belongs to the coordinated users if and only if $\mathcal{l}_v = 1$. 
To keep us general, we say that $v$ is a {\it $1-$node} if $\mathcal{l}_v = 1$  and a {\it $0-$node} otherwise. We collect the labels in the vector $\mathcal{l}$.

We denote with $N (v)$ the out-neighbourhood of a node $v$ (without $v$ itself) and as $d(v)$ its out-degree. Moreover, we define $d_0^{\mathcal{l}}(v) = \{w \in N (v) \; | \; \mathcal{l}_w = 0\}$, i.e. the number of $0-$nodes in the out-neighbourhood of $v$. A similar definition holds for $d_1 ^{\mathcal{l}}(v)$.

Given a labelling $\mathcal{l}$, we denote the subset of $1-$nodes in $V$ by $V_1 ^\mathcal{l} = \{ v \in V \; | \; \mathcal{l}_v = 1\}$, 
and its cardinality by $k_{\mathcal{l}} = \lvert V_1 ^\mathcal{l} \rvert$, i.e. the number of $1-$nodes in $V$.

Finally, $p_{\mathcal{l}} (w)$ is the binary variable indicating if the parent of $w$ is a $1-$node (i.e. it has a direct interaction with a coordinated account).

We define the {\it influence} of a configuration $(T,\mathcal{l})$ as

\begin{equation}\label{eq:influence}
    I_{\mathcal{l}}(T) = \sum_{v \in V_1 ^\mathcal{l}} d_0^{\mathcal{l}} (v).
\end{equation}

Equation \eqref{eq:influence} counts the number of non-coordinated users that are connected (i.e. share content) of a coordinated one. Therefore, it provides a measure of the impact that coordinates users exert on non-coordinated ones.

We are interested in defining an algorithm that searches the labelling (i.e. the disposition of coordinated accounts) that maximises \eqref{eq:influence}. We denote the optimum influence among all the labelings by $I_{\mathcal{l}}^*(T)$ and its labeling using the minimum number of coordinated accounts by $\mathcal{l}^*(T)$.
Accordingly, we denote as $V_1 ^*$ and $k^*(T) = \lvert V_1 ^*\rvert$ the set of optimal $1-$nodes and its cardinality.
We highlight that, when the context is clear, we will omit $\mathcal{l}$ and $T$ from all the previous notations.

\begin{figure}[!ht]
    \centering
    \includegraphics{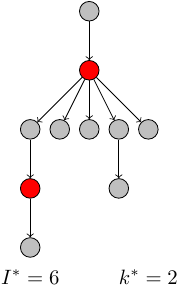}
    \caption{Example of an optimal configuration. The coordinated accounts are highlighted in red.}
    \label{fig:example_definitions}
\end{figure}

For example, consider the tree depicted in Figure \ref{fig:example_definitions}. Among all the possible labels, the optimal one has $I^* = 6$ and $k^* = 2$. Indeed, even for any labelling of size $k=3$, the influence reaches at most $6$.

\begin{figure}[!ht]
    \centering
    \includegraphics{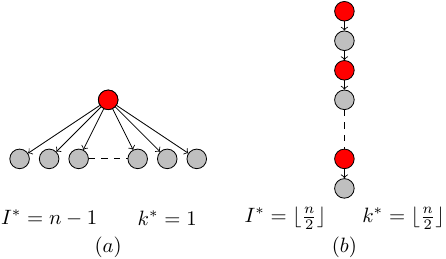}
    \caption{$(a)$ The configuration that maximizes $I^* (n)$ and minimizes $k^* (n)$. $(b)$ The configuration that minimizes $I^* (n)$ and maximizes $k^* (n)$.}
    \label{fig:bounds}
\end{figure}

Interestingly, it is possible to give some constraints over the possible values of $I^*$ and $k^*$. In more detail, let $T_n$ be the set of all directed trees with $n$ vertices. Then, $\max\{ I^* (T):\; T\in T_n\}=n-1$ and it is obtained for the configuration depicted in Figure \ref{fig:bounds} $(a)$. Note that it corresponds also to the lower bound for $k^*$. On the other hand, $\min\{ I^* (T):\; T\in T_n\}= \lfloor \frac{n}{2} \rfloor$ and it is obtained for the configuration depicted in  Figure \ref{fig:bounds} $(b)$ which also provides the upper bound for $k^*$.

Therefore, for a general directed tree $T$ on $n\geq 2$ vertices, the following inequalities hold: 

\begin{equation}\label{eq:boundI}
    \lfloor \frac{n}{2} \rfloor \leq I^* (T) \leq n-1,
\end{equation}

\begin{equation}\label{eq:boundk}
    1 \leq k^* (T) \leq \lfloor \frac{n}{2} \rfloor.
\end{equation}

\section{Data and methods}\label{sec:data_methods}

\subsection{Data}\label{sec:data_collection}

Our dataset for this study is based on a collection of tweets related to the online debate about the 2019 United Kingdom general elections presented in~\cite{nizzoli2021coordinated}.
In brief, data were collected using official Twitter API and performing both hashtag search (\#GeneralElection19, \#VoteLabour, \#VoteConservative etc.) and timeline download of the involved political parties and their leaders. The time window for the download lasted for one month, from 12 November to 12 December 2019, and the volume of downloaded data is 11,264,820 tweets published by 1,179,659 distinct users. The dataset is available at \url{https://doi.org/10.5281/zenodo.4647893}. The set of coordinated accounts is also the same provided in~\cite{nizzoli2021coordinated} in which a method for coordinated account detection is presented. The method is made up of the following steps: 1) Selection of influential users or super-spreaders; 2) Selection of users similarity measure (e.g. cosine similarity); 3) Creation of a users similarity network using pairwise comparison; 4) Filtering of users similarity network; 5) Clustering of the similarity network. This method provides clusters of users provided with a continuous coordination score. 
The distinction between coordinated and non-coordinated accounts and the reconstruction of information cascades was performed in a later work~\cite{cinelli2022coordinated} that we describe in the following lines.
Coordinated accounts were selected by retaining only the nodes at the endpoints of the 1\% of the links having the highest similarity scores. %Considering such binary labelling, a further filtering step was performed in order to retain tweets that were retweeted at least once by the accounts classified as coordinated and whose author is not a coordinated user. 
The number of tweets is 49,331, and for each of them an information cascade based on retweets was reconstructed using a method proposed in~\cite{de2015towards}. The motivation for using a specific method for cascade reconstruction is due to the fact that Twitter API didn't (and do not) provide the full structure of the retweet trees so that if a user $j$ retweets $t$ by user $i$ and user $k$ retweets the same tweet $t$ from the retweet of user $j$ the edge list provided by the Twitter API will be ($i \rightarrow j$; $i \rightarrow k$), thus a star. However, this star is not a proper representation of the actual information flow, being it ($i \rightarrow j$; $j \rightarrow k$). For this reason, the method proposed in~\cite{de2015towards} provides a deterministic way for reconstructing retweet-based information cascades that relies on the knowledge of friendship networks (i.e. following relationships) and on the assumption that a user will be retweeting a tweet from the latest retweeters of $t$ contained in her/his list of friends. Therefore, assuming the same situation as before, but knowing that user $k$ follows user $j$, the reconstructed cascade will be ($i \rightarrow j$; $j \rightarrow k$). Using this procedure on the selected 49,331 tweets and their retweets the same number of information cascades was reconstructed. Details about the structure of such cascades are provided in Table~\ref{tab:twitter_trees}.

\begin{table}[!ht]
    \centering
    \begin{tabular}{|c|c|}
       \hline
       Number of cascades & 49331 \\
       \hline 
       Minimum number of nodes & 1 \\
       \hline 
       Maximum number of nodes & 9066 \\
       \hline
       Minimum number of coordinated accounts & 0 \\
       \hline
       Maximum number of coordinated accounts & 236 \\
       \hline
    \end{tabular}
    \caption{Data breakdown of the cascades collected from Twitter.}
    \label{tab:twitter_trees}
\end{table}

\subsection{Optimal placement of coordinated accounts without constraints on their number}

In this section we design an algorithm that, starting from a directed tree $T$ rooted at node $r$, provides $I^*(T)$ and $V_1^*(T)$. We highlight that, in this section, we are not imposing an upper bound to the number of $1-$nodes, resembling a case of unlimited resource capability. 

For a node $v$, let $\mathrm{MI}(v)$ denote the value of the maximum influence in the subtree of $T$ rooted at $v$, and $V_1$ denote the set of the coordinated accounts of $T$. Moreover, let $\mathrm{MIyes}(v)$ and $\mathrm{MIno}(v)$ be the values of the maximum influence of the subtree rooted in $v$ that includes $v$ in $V_1$, and that excludes $v$ from $V_1$, respectively.  
Our aim is to compute $\mathrm{MI}(r)=\max\{ \mathrm{MIyes}(r), \; \mathrm{MIno}(r)\}$. 

We give a recursive definition of $\mathrm{MIyes}(r)$ and $\mathrm{MIno}(r)$. The base cases are obtained when the height of the tree is zero or one.
If the height is zero, it means that $V = \{r\}$ and $E = \emptyset$. In this case, $\mathrm{MIyes}(r)=\mathrm{MIno}(r)=0$, and $\mathrm{MI}(r)=\mathrm{MIno}(r)=0$ since the maximum influence is 0.

If the height is one and $V = \{r=v_1,v_2,\ldots v_n\}$ then, without loss of generality, let $E = \{(v_1,v_2),\ldots (v_1,v_n)\}$ (i.e. $T$ is the tree depicted in Figure \ref{fig:bounds}a). Then,  
$\mathrm{MIyes}(r)=\sum_{w\in N(r)} (\mathrm{MIno}(w)+1)$, since if $r\in V_1$ the maximum influence is equal to the sum of the children nodes not belonging to $V_1$.
In this case $\mathrm{MIno}(w)=0$ for all $w\in N(r)$ being leaves, and so we get 
$$\mathrm{MI}(r)= \mathrm{MIyes}(r)=\sum_{w\in N(r)} 1 = d(r).$$

Suppose $T$ has height more than one. The maximum influence is computed by considering the maximum influence of the subtrees rooted at the children of $r$. Let's denote the children of $r$ by $w_1,\ldots, w_{d (r)}$, and $T_1 \ldots T_{d (r)}$ the subtrees rooted at each child of $r$. 
Then, the following recurrence relation describes how to obtain $\mathrm{MI}(r)$:

\begin{equation}
    \mathrm{MI}(r)=\max 
    \begin{cases}
        \mathrm{MIno}(r) =\sum_{w\in N(r)} \max(\mathrm{MIyes}(w),\;\mathrm{MIno}(w) )\\
        \mathrm{MIyes}(r) =\sum_{w\in N(r)} \max(\mathrm{MIyes}(w),\mathrm{MIno}(w)+1)  
    \end{cases}
\end{equation}

Indeed, if the maximum influence is obtained by $r\notin V_1$, then it is the sum of the values of the maximum influences for the subtrees rooted at its children. Otherwise, if it is obtained for $r\in V_1$, it is the sum of the values of the maximum influences for the subtrees rooted at its children and adding one for every child that is not a coordinated node. 

We memoize the functions  $\mathrm{MIyes}$ and $\mathrm{MIno}$ into the tree itself by defining two fields for each node $v$, that is $v.\mathrm{MIyes}$ and $v.\mathrm{MIno}$. The algorithm computes the label of each node $v.\mathcal{l}$ by comparing $v.\mathrm{MIyes}$ and $v.\mathrm{MIno}$.
Node $v$ is included in $V_1$ if and only if $v.MIyes > v.MIno$ (i.e. if $d_0(v) > 0$). 

The implementation uses a post-order tree traversal realised by one recursive call. We denote the label obtained by the algorithm as $\bar{\mathcal{l}}$.

\begin{algorithm}[!ht]
\caption{$\mathrm{TreeMaxInfluence}$}
\label{alg:recursiveOneCall}
\begin{algorithmic}
        \Require $v$
        \Ensure $I^*(T_v), V_1^{\bar{\mathcal{l}}}$
    \State $v.\mathrm{MIno}\gets 0$
     \State $v.\mathrm{MIyes}\gets 0$
   \If{$v$ is a leaf} 
        \State {$v.\mathcal{l}\gets 0$} \Comment{$v\notin V_1$}
       \State {\bf return} $v.\mathrm{MIno}$ \Comment{the Max Influence value is zero}
   \Else \Comment{It has at least one child}

     \For {each child $w$ of $v$}  
        \State $\mathrm{treeMax} \gets \mathrm{TreeMaxInfluence}(w)$ 
        \State $v.\mathrm{MIno}\gets  v.\mathrm{MIno}+\mathrm{treeMax}$ \Comment{Compute the field $\mathrm{MIno}$ of $v$}
        \If{$\mathrm{treeMax} = w.\mathrm{MIno}$}
             \State{Increment $\mathrm{count}$} \Comment{Count the children $w\notin V_1$}
        \EndIf

       \EndFor
       \State $v.\mathrm{MIyes}\gets v.\mathrm{MIno}+ \mathrm{count}$ \Comment{Compute the field $\mathrm{MIyes}$ of $v$}
       
       \If{$v.\mathrm{MIyes} > v.\mathrm{MIno}$}
            \State{$v.\mathcal{l}\gets 1$} \Comment{$v\in V_1$}
            \State {\bf return} $v.\mathrm{MIyes}$ \Comment{Return the Max Influence value}
        \Else
            \State{$v.\mathcal{l}\gets 0$} \Comment{$v\notin V_1$}
            \State {\bf return} $v.\mathrm{MIno}$ \Comment{Return the Max Influence value}
        \EndIf
    \EndIf
\end{algorithmic}
\end{algorithm}

\begin{Thm}
$TreeMaxInfluence(r)$ computes the optimal value of the influence of $T$ in $O(|V|)$ time.
\end{Thm}

\proof

Suppose to start with an optimal labelling of each of the subtrees rooted at the children $w_i$ of $r$, i.e. $T_i$, with $i = 1,\ldots d(r)$. 
If $r.\mathrm{MIyes} \leq  r.\mathrm{MIno}$ then $r$ is not inserted in $V_1$ and the configuration is optimal. 

Otherwise, if $r.\mathrm{MIyes} > r.\mathrm{MIno}$, $r$ is included into $V_1$. If all the $r$-children are $0$-nodes, $r.\mathrm{MIyes}$ is the optimum. 
Suppose that $\mathrm{MI}(r)> r.\mathrm{MIyes}$, that is, it is not optimal and let $\mathcal{l}$ be the computed labelling. Notice that since $r.\mathrm{MIyes} > r.\mathrm{MIno}$ and $r.\mathrm{MIyes}= r.\mathrm{MIno}+ d_0 (r)$, then $d_0 (r)>0$. 
 Since $r.MIyes$ is not optimal, there are at least two nodes $w_i,w_j$ such that $\mathcal{l}(w_i)=0$, and $\mathcal{l}(w_j)=1$. 
We have that $w_j.\mathrm{MIyes} > w_j.\mathrm{MIno}$, where $w_j.\mathrm{MIyes}= w_j.\mathrm{MIno}+ d_0 (w_j)$. Thus, $d_0 (w_j)>0$ which implies  $w_j.\mathrm{MIyes}-1\geq  w_j.\mathrm{MIno}$. 
To increase the influence in the tree rooted at $r$ with $r\in V_1$, one more child should be added to the $0$-nodes. 
Therefore let $\bar{\mathcal{l}}$ be the new labelling that differ only for $\bar{\mathcal{l}}(w_j)=0$. 
This leads to a change of
\begin{equation}
    \Delta (I) = 1 - w_j.\mathrm{MIno}-w_j.\mathrm{MIyes} = d_0 (r)+1- d_0 (r) + w_j.\mathrm{MIno}-w_j.\mathrm{MIyes}
\end{equation}
since all the other terms cancel each others (they are unchanged in the two labellings). Moreover, by $w_j.\mathrm{MIyes}-1\geq  w_j.\mathrm{MIno}$ follows that $\Delta (I) \leq 1+ w_j.\mathrm{MIyes}-1 -w_j.\mathrm{MIyes}\leq 0$. This means that the influence cannot be increased, and hence it was maximal.

To conclude, since each node is visited once and some constant-time operations are applied in each case, the algorithm runs in $O(|V|)$ time.
\qed

% Suppose to start with an optimal labelling of each of the subtrees rooted in the children of $r$, i.e. $T_1, T_2 ,\ldots, T_{d (r)}$.
% If $r.MIyes \leq r.MIno$ then $r$ is not inserted in $V_1$ and the configuration is optimal.

% Otherwise, if $r.MIyes > r.MIno$, $r$ is included into $V_1$. 
% If the result is not optimal, then the labelling of at least one of the $T_i, i = 1,\ldots d(r)$, such that $\mathcal{l}(w_i) = 1$, can be modified to improve $I$.

% Let $T_i$ be the tree rooted in node $w_i \in N (r)$, where $\mathcal{l}(w_i) = 1$, and let $v \in V(T_i) \setminus V_1 (T_i)$. We impose $\mathcal{l}_{w_i} = 0$ and $\mathcal{l}_v = 1$.
% This led to a change of 

% \begin{equation}
%     \Delta (I) = d_0 (r) + 1 - d_0  (w_i) + d_0 (v).
% \end{equation} 

% Recall that both $d_0 (r)$ and $d_0  (w_i)$ must be strictly positive quantities, otherwise the algorithm does not include them in $V_1$). 
% It's easy to see that, to obtain $\Delta (I) > 0$, $d_0 (v)$ must be at least greater than $0$ (but other worst cases may appear depending on the value of $d_0 (r) + 1 - d_0  (w_i)$). 
% However, this is not possible since $v$ was not in $V_1(T_i)$ before the label update.

% To conclude, since each node is visited once and some constant-time operations are applied in each case, the algorithm runs in $O(|V|)$ time.
% \qed

Although Algorithm \ref{alg:recursiveOneCall} returns the optimal value of the influence, it is possible that $ \lvert V_1^{\bar{\mathcal{l}}} \rvert \geq k^*$. To solve this problem and obtain $\mathcal{l}^*$, we use Algorithm \ref{alg:clear1nodes}.

\begin{algorithm}[!ht]
\caption{$clear1nodes$}
\label{alg:clear1nodes}
    \begin{algorithmic}
        \Require $T = (V,E)$, $V_1 ^{\bar{\mathcal{l}}}$
        \Ensure A minimum 1-labelling $\mathcal{l}^*$

        \If{$d_0 ^\mathcal{l}(r) = 0$ and $d_0 ^\mathcal{l}(w_i) = 1$ for each $w_i \in N (r)$}
            \State $r.\mathcal{l} = 1$
            \State $w_i .\mathcal{l} = 0$ for $w_i \in N(r)$
        \EndIf
        \For {each $v \in V_1 ^{\bar{\mathcal{l}}}$} 

            \If{$d_0^\mathcal{l}(v)  = p_{\mathcal{l}}(v)$}

            \State $v.\mathcal{l} \gets 0$ \Comment{$v\notin V_1$ }

            \EndIf

        \EndFor
    \end{algorithmic}
\end{algorithm}

The correctness of the algorithm easily follows: 
if all the children of $r$ are $1-$nodes and their out-degree is $1$, we can simply put $r$ in the $1$-nodes and remove all its children. This procedure reduces the number of used coordinated accounts without changing the influence.
Moreover, since each node $v \in T$ has exactly one father (except for the root), the only case in which $v$ does not contribute to the influence is when $d_0 (v) = 1$ and its father is a $1-$node. The algorithm runs in $O(k^*)$.

\begin{figure}[!ht]
    \centering
    \includegraphics{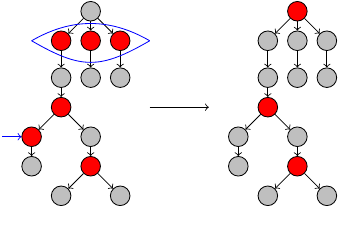}
    \caption{$(a)$ a directed tree and the label obtained trough Algorithm \ref{alg:recursiveOneCall}. The $1-$nodes highlighted by the blue arrows can be removed. $(b)$ After a run of Algorithm \ref{alg:clear1nodes} the nodes are deleted and $V_1^*$ is returned.}
    \label{fig:clear1nodes}
\end{figure}

A visual example of the running of Algorithm \ref{alg:clear1nodes} is depicted in Figure \ref{fig:clear1nodes}.

\subsection{A greedy placement with constraints on the number of coordinated accounts}

In this section we propose a greedy strategy to face the case in which we want to maximize $I$ with a fixed number $k$ of coordinated accounts. 

We define the {\it switch(v,w)} operator that, starting from a labelling $\mathcal{l}$, provides a new labelling $\bar{\mathcal{l}}$ that increase the obtained influence:

\begin{Def}
Let $\mathcal{l}$ be a labelling of nodes of a given directed tree $T$.
Suppose $v \in V_1^l$ and $w \in V \setminus V_1 ^\mathcal{l}$; the switching operator $switch(v,w)$ exchanges the labels between $v$ and $w$, i.e. it sets $\mathcal{l}_v = 0$ and $\mathcal{l}_w = 1$. 
\end{Def}

Thus, the operator provides a new labelling $\bar{\mathcal{l}}$, where $V_1^{\bar{\mathcal{l}}} = \left( V_1^{\mathcal{l}} \setminus \{v\} \right) \cup \{w \}$. 

Based on this idea, we apply the following strategy: we start by including sequentially that of the $k$ nodes which contribute the most to the influence. Then, we try to increment the influence by applying {\it trySwitch}. When it is no longer possible to increase $I$ using a call of {\it switch}, the algorithm stops.

\begin{figure}[!ht]
    \centering
    \includegraphics{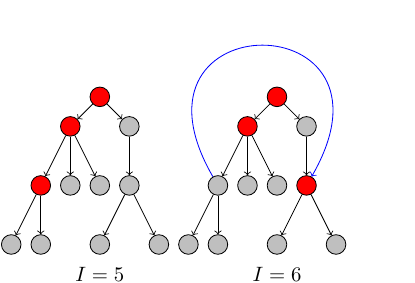}
    \caption{An example of how {\it switch} is applied.}
    \label{fig:switch}
\end{figure}

To formally define the concept of increment, consider a node $v \in V_1^\mathcal{l}$ and $w \in V \setminus V_1^\mathcal{l}$. 
Switching $l(v$) with $l(w)$ leads to an increment if

\begin{equation}\label{eq:increment}
d^{\mathcal{l}}_0 (v) < \Delta(v,w) := d^{\bar{\mathcal{l}}}_0 (w) - p_{\bar{\mathcal{l}}} (w) + p_{\bar{\mathcal{l}}}(v),
\end{equation}

where $\bar{\mathcal{l}}$ is the label resulting from $switch(v,w)$.

The motivation is easily obtained: $d^{\bar{\mathcal{l}}}_0 (w)$ counts the number of $0-$nodes in the out-neighbour of $w$.
Moreover, when changing $v$ with $w$ it is possible to add $1$ to the influence if the parent of $v$ is a $1-$node, i.e. $p_{\bar{\mathcal{l}}}(v)=1$. Similarly, we subtract $1$ from the influence if the parent of $w$ is a $1-$node i.e. $p_{\bar{\mathcal{l}}} (w)=1$.

Summarizing, if $\eqref{eq:increment}$ is satisfied, $switch(v,w)$ assures that $I_{\mathcal{l}} > I_{\mathcal{l}}$.

Algorithm \ref{alg:bot_placement_greedy} implements the pseudo-code of the greedy strategy.

\begin{algorithm}[!ht]
\caption{{\it trySwitch}}
\begin{algorithmic}
\Require $T = (V,E), \mathcal{l}$
\Ensure $\Bar{\mathcal{l}}$
\State $check = TRUE$
\While{$check$}\Comment{Try to apply {\it switch}}
    \State $check = FALSE$
    \For{$v \in V_1^\mathcal{l}$}
      \If{$d_0 (v) < \max_{u \in V \setminus V_1^\mathcal{l}} \Delta(v,u)$} \Comment{It is possible to increase the influence}
      \State $check = TRUE$
      \State $S = \{ w \in V \setminus V_1^\mathcal{l} \; | \; \Delta(v,w) = \max_{u \in V \setminus V_1^\mathcal{l}} \Delta(v,u)\}$
      \State choose randomly $u \in S$
      \State $\mathcal{l} = switch(v,u)$
      \EndIf
      \EndFor
\EndWhile
\State $\Bar{\mathcal{l}} = \mathcal{l}$
\end{algorithmic}

\end{algorithm}

\begin{algorithm}
\caption{An heuristic for the optimal bot placement}\label{alg:bot_placement_greedy}
\begin{algorithmic}
\Require $T = (V,E)$, $\mathcal{l}$,$k$
\Ensure an approximation $\Tilde{\mathcal{l}}$ of the optimal solution

\For{$i = 1:k$}\Comment{Place the first $k$ nodes} 

     \State $W = \{ w \in V \setminus V_1^\mathcal{l} \; | \; d_0(w) - p_{\mathcal{l}}(w) = max_{u \in V} \left( d_0 (u) - p_{\mathcal{l}} (u) \right)$\}
     \State Select randomly a node $v \in W$ 
     \State $\mathcal{l}(v) = 1$
\EndFor
\State $\Tilde{\mathcal{l}} = trySwitch(T,\mathcal{l})$
\end{algorithmic}
\end{algorithm}

\section{Results}
\subsection{Growth rate of $I^*$ and $k^*$}

In this section we are interested in determining the growth rate of $I^*$ and $k^*$ depending on the number of nodes (i.e. users) and on height of the tree.
Although those values are bounded, as shown in \eqref{eq:boundI}-\eqref{eq:boundk}, finding an explicit expression for them it's beyond the scope of this work.

\begin{figure}[!ht]
    \centering
    \includegraphics[width = \linewidth]{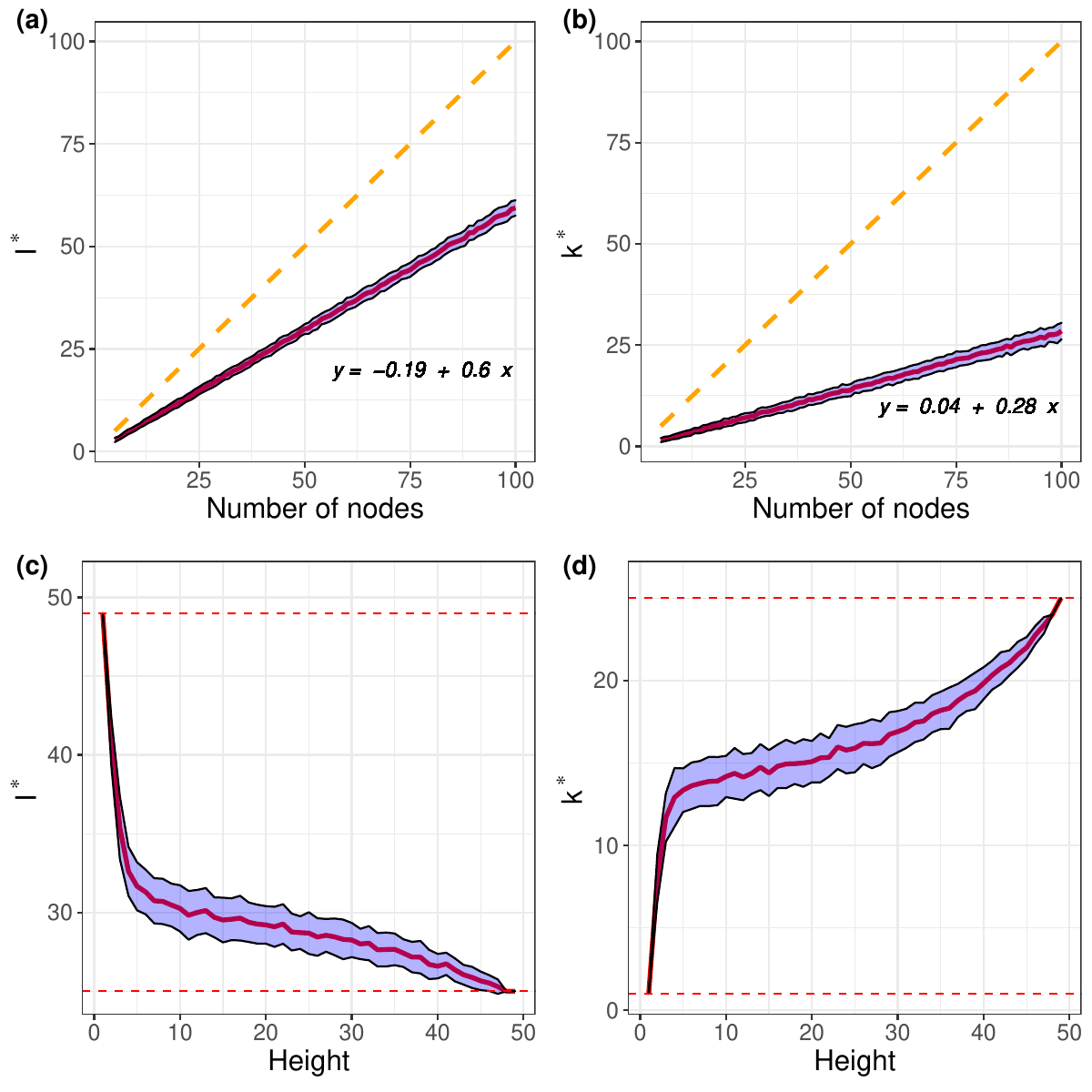}
    \caption{Upper row: relation between the number of nodes of a tree and $(a)$ mean optimal influence $(b)$ mean optimal number of bots. 
    Bottom row: comparison between the height of the tree and $(c)$ mean optimal influence and $(d)$ mean optimal number of nodes. 
    The two horizontal red lines show the maximum and minimum values of the measure, while the ribbon around the curves represents a standard deviation from the mean.}
    \label{fig:bot_simulation}
\end{figure}

To obtain an approximation we consider $N = 100$ random trees made up of $5 \leq n \leq 100$  nodes. We apply to each of them Algorithm \ref{alg:recursiveOneCall} and Algorithm \ref{alg:clear1nodes} to obtain $I^*$ and $k^*$. 
Then, we average the results over the $N$ graph realisations for each of the $n$ values We plot the results in Figure \ref{fig:bot_simulation}$(a)$ and $(b)$.

Interestingly, we observe in both cases a linear relation ($p < 0.001$, $R^2 = 0.99$) in which the slope is much lower than $1$ ($\approx 0.28)$ in the case of $k^*$ while closer to the unity $(\approx 0.6)$ in the case of $I^*$.
This suggests that it is possible to reach the point of maximum influence using approximately $0.3\cdot n$ coordinated accounts. Moreover, they are sufficient to exert a great influence on the tree (approximately $0.6\cdot n$).

To evaluate the impact of the tree's height on $I^*$ and $k^*$, we consider several random directed trees having a fixed number of nodes ($n = 50$), and height spanning between $1$ and $n$. 
For each $h$, we generate $N = 100$ random trees having height $h$. Similarly to the previous case, we then average the results obtained by our algorithms, which are shown in Figure \ref{fig:bot_simulation}$(c)$ and $(d)$.
As depicted, $I^*$ drops dramatically as soon as the height increases, and starts to decrease more softly when $h \geq 10$. A symmetrical result holds for $k^*$.
This suggests an expected result, i.e. for a fixed number of users in the conversation, information cascades in which the majority of the retweets are close to the original user (i.e. the associated tree has a low height) can be covered with fewer coordinated accounts, obtaining a greater impact.

However, the Figure also suggests that from a certain height threshold ($\approx 20\cdot n$ in our simulation case) $I^*$ starts to show greater stability with respect to the increase in the height of the tree.

Symmetrical results hold for $k^*$.

\subsection{Phase diagram of the relative positions of node labels}

In Section \ref{sec:data_methods} we defined an algorithm that finds one of the optimal placements node labels. Interestingly, the same problem can be translated into finding an assignment of binary node metadata such that the number of (directed) edges between $1-$nodes and $0-$nodes is maximised~\cite{cinelli2020network}. 

In such a framework, let's denote with $m_{11}$ ($m_{10}$) the number of directed edges between $1$ and $1$ ($0$).
To better understand how rare the optimal placement is when compared to the set of all the labellings, we computed the phase diagram relative to $m_{11}$ and $m_{10}$~\cite{park2007distribution}. 

\begin{figure}[!ht]
    \centering
    \includegraphics[scale = 0.7]{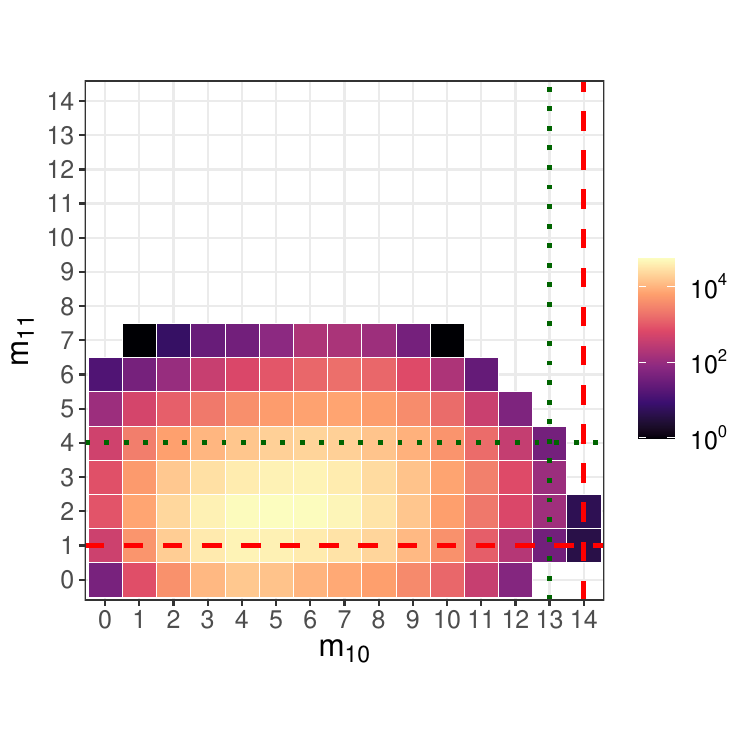}
    \caption{Phase diagram of possible values of $(m_{10},m_{11})$. The intersection of red (green) lines highlights $(m_{10},m_{11})$ for the optimal (greedy) placement.
    The colours that fill the bins use a logarithmic scale.}
    \label{fig:phase_diagram}
\end{figure}

More in detail, we generate a random directed tree of $n = 25$ nodes and compute its optimal influence, obtaining $I^* = 14, k^* = 7$. Then, we generated all the $\binom{25}{7}$ possible placements, storing for each of them the corresponding values of $m_{11}$ and $m_{10}$. 
We show a $2d-$histogram of the results in Figure \ref{fig:phase_diagram}.

We observe that both the optimal and greedy placements are located in low-frequency areas at the boundary of the diagrams, obtaining high values of $m_{10}$. This provides results related to the statistical significance of the obtained labellings and suggests that they are somewhat rare with respect to the whole sets of labels. In general, a placement without a specific strategy is less likely to obtain good results. 

\subsection{Results of the algorithm on real cascades}

In this section we consider a dataset of Twitter's cascades (modelled as directed trees) in which nodes are labelled as either coordinated or non-coordinated as explained in section \ref{sec:data_collection}. In each cascade, we aim to detect the influence exerted compared to the optimum placement provided by Algorithm \ref{alg:recursiveOneCall}. 
In particular, the {\it size} of a cascade is the number of its nodes (i.e. users).

For a given cascade $c$, we denote as $I(c)$ the influence obtained by them and as $k(c)$ the number of coordinated accounts acting in $c$. The optimal influence and number of coordinated accounts are denoted accordingly as $I^*(c)$ and $k^*(c)$.

For our analysis, we only consider cascades having at least $15$ nodes and at least $1$ coordinated account. Such a restriction allows us to consider more complex properties and structures at the cost of losing some of the data.
After this procedure, we obtain 4119 trees to which we apply Algorithm \ref{alg:recursiveOneCall} and Algorithm \ref{alg:clear1nodes} to compute $I^*(c)$ and $k^*(c)$.

In each cascade we measure how close $I(c)$ is to $I^*(c)$ using 

\begin{equation}\label{eq:influence_normalized_with_optimum}
\rho(c) = \frac{I(c)}{I^*(c)}.
\end{equation}

Note that $0 \leq \rho(c) \leq 1$, where a value close to one indicates that the influence obtained by the detected coordinated accounts is close to the optimal value.

\begin{figure}[!ht]
    \centering
    \includegraphics[scale = 0.8]{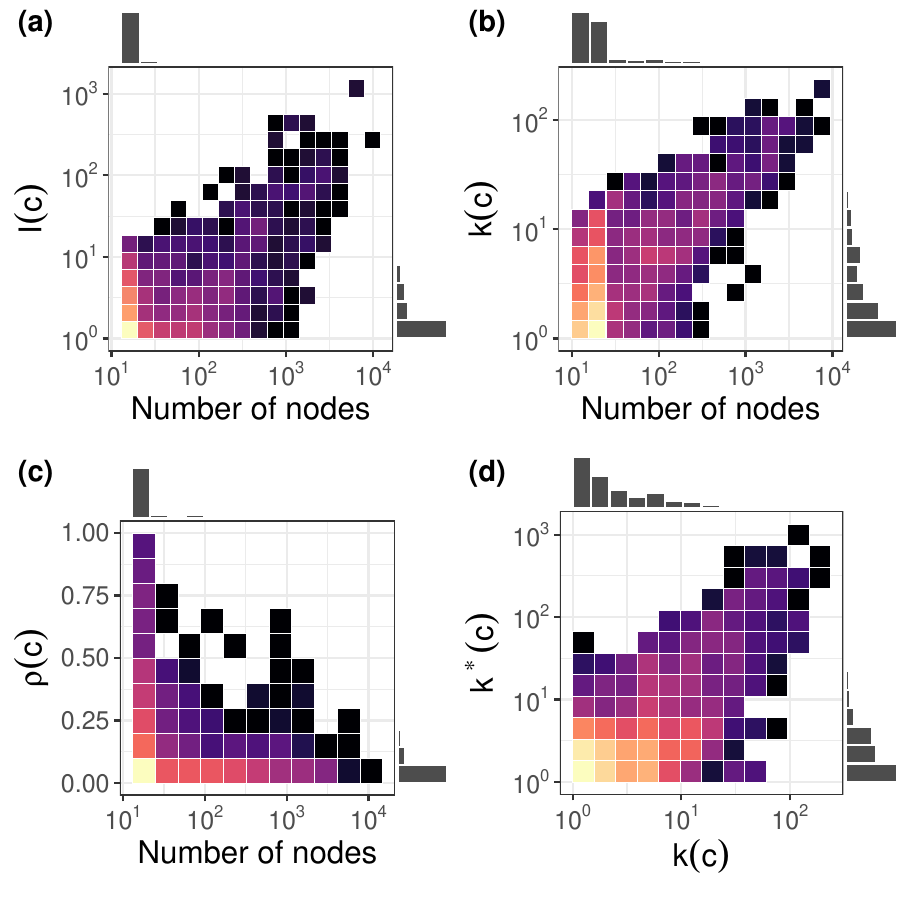}
    \caption{Results of the algorithm on real cascades. $(a)$ and $(b)$ show the $2$-dimensional density of the number of nodes and $(a)$ $I(c)$ or $(b)$ $k(c)$. In $(a)$ we add 1 to the influence of each tree for graphical reason. $(c)$ $2-$dimensional density of the number of nodes and $\rho(c)$. $(d)$ $2-$dimensional density of $k(c)$ and $k^*(c)$.
    The colours that fill the bins use a logarithmic scale.}
    \label{fig:results_real_networks}
\end{figure}

As reported in Figure \ref{fig:results_real_networks}, the majority of the cascades have a limited size and, in general, the influence obtained from the coordinated accounts is low, as depicted in $(a)$. In fact, in the biggest trees, $I(c)$ is approximately $10\%$ of the number of nodes, much lower than the expected $60\%$ suggested by Figure \ref{fig:bot_simulation}$(a)$. A somewhat similar result is observed in $(b)$, which indicates the majority of the cascades contain a very low number of coordinated accounts. Also in this case, approximately $1\%$ of the nodes is coordinated in the biggest cascades, while Figure \ref{fig:bot_simulation}$(b)$ suggests that $30\%$ of the number of nodes is needed to obtain the optimal results. Accordingly, Figure \ref{fig:results_real_networks}$(c)$ shows that only small cascades succeed in obtaining influence values close to the optimal one. For larger cascades, we observe a dramatic collapse in $\rho(c)$, probably due also to the much more complex structures arising with a higher number of nodes. 
Finally, $(d)$ shows that, apart from the smallest trees in which the two measures are approximately similar, $k(c)$ largely differ from $k^*(c)$, yielding a possible explanation of the low values of $\rho(c)$.

Summarising, we conclude that coordinated accounts exert an influence much lower than the upper bound provided by our algorithm. Moreover, there is evidence that this may be due to a limited number of coordinated accounts placed in each cascade.

\subsection{Comparison with optimum using a fixed number of coordinated accounts}

In the previous section we compared the behaviour of coordinated accounts with an optimal strategy that considers having, at prior, an undefined number of coordinated accounts to place.
As highlighted before, the scarce results obtained by them could be due to a limited number of resources employed to influence other nodes.

Therefore, here we are interested in unveiling if the observed non-optimal behaviour changes if we compare it to the greedy strategy proposed in Algorithm \ref{alg:bot_placement_greedy}, using exactly $k(c)$ coordinated accounts. 
Observing poor results in this case could suggest that they are due also to a bad placement strategy of coordinated accounts.

We denote as $I_k(c)$ the influence computed by Algorithm \ref{alg:bot_placement_greedy}. Similarly as before, we measure how close $I(c)$ and $I_k(c)$ are using

\begin{equation}
    \rho_k (c) = \frac{I(c)}{I_k(c)}.
\end{equation} 

\begin{figure}[!ht]
    \centering
    \includegraphics[scale = 0.7]{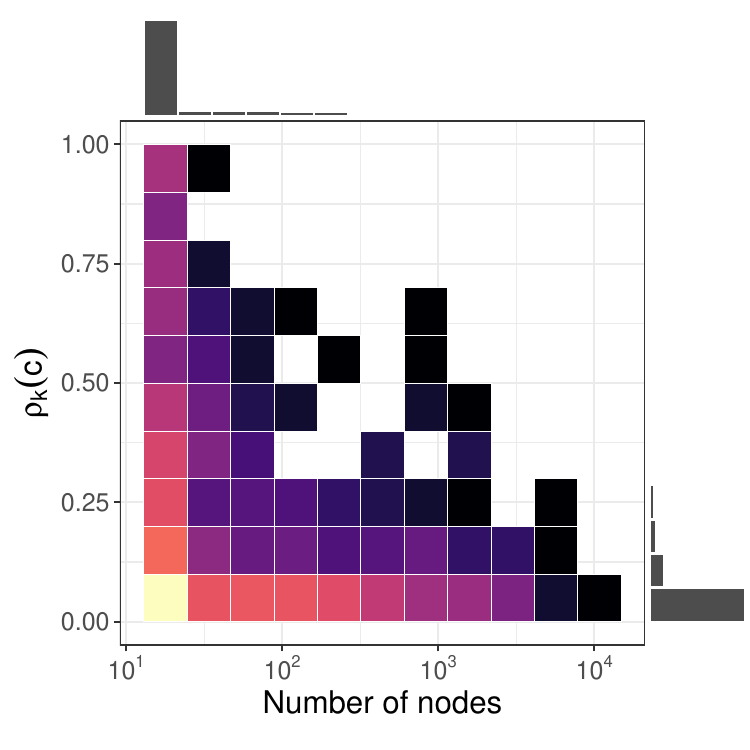}
    \caption{$2-$dimensional density of the number of nodes and $\rho(c)_k(c)$. The colours used to fill the bins use a logarithmic scale.}
    \label{fig:greedy_strategy}
\end{figure}

In Figure \ref{fig:greedy_strategy} we show a $2-$dimensional histogram of $\rho_k(c)$ versus the number of nodes in the cascade.
Notably, the figure looks very similar to Figure \ref{fig:bot_simulation}(c): only small cascades get comparable results with the greedy algorithm but, when the size of the cascade increases, the placement of coordinated accounts in real cascades hardly reach the influence obtained through the greedy strategy.

Taking into account all the previous results, it seems that, in real cascades, coordinated accounts may be placed randomly, without a specific rationale.
To validate this observation, for each cascade we generate $N = 10$ random labellings and we compute the mean influence obtained by those random placements.
This results in a new distribution of influence values, denoted as $q_{random}$. 
We also denote as $p_{real},q_{greedy}$ the real and greedy influence distributions.

We then compute Kullback-Leibler divergence \cite{kld} to compare $p_{real}$ with the other two distributions, adding a small correction of $10^{-4}$ to handle the cases in which $q_{greedy}$ and $q_{random}$ distributions have $0$ probability. 
The results are shown in Table \ref{tab:kld}.

\begin{table}[!ht]
    \centering
    \begin{tabular}{|c|c|}
    \hline
        $D_{KL}(p_{real}||q_{greedy})$ & $4.278$ \\
        \hline
        $D_{KL}(p_{real}||q_{random})$ & $0.097$ \\
        \hline
    \end{tabular}
    \caption{Kullback-Leibler divergence values.}
    \label{tab:kld}
\end{table}

The values suggest that, as expected, $p_{real}$ is much more similar to $q_{random}$ than $q_{greedy}$, thus confirming that, according to our model, coordinated nodes are placed quite randomly.

Joining these results with the previous ones, we can conclude the results observed in real cascades are scarce for two reasons: a scarcity of resources and also a bad placement strategy.

\section{Conclusion}
In this work we have proposed a general framework that allows to evaluate the influence of coordinated accounts in real cascades, providing an upper bound to its value. 

First, we show (using synthetic data) that it is possible to exert the maximum influence with a low number of coordinated accounts. Moreover, the majority of node labelling gives far-from-optimum results, confirming that random placement is not likely to obtain good results.

As a case of study, we consider $\approx 4K$ information cascades on Twitter about $2019$ UK political elections.
We show that observed coordinated accounts exert a very low influence on the tree, compared to both greedy and optimal strategies. 
Our results suggest that this is due to a double reason: a scarcity of resources (i.e. used coordinated accounts) and an absence of a strategy in how coordinated accounts act. 

Despite the possible limitations of our model and the limitations of the detection procedure and the reconstruction of Twitter cascades, our results overall suggest that CIB may play a less pivotal role than expected in information diffusion. Nonetheless, as often reported in social media studies~\cite{ruths2019misinformation}, different methodological settings and case studies may lead to different results thus future work may employ the proposed framework to consider the impact of CIB and other similar actors in a wider range of scenarios.

\section*{Acknowledgement}
The work is supported by IRIS Infodemic Coalition (UK government, grant no. SCH-00001-3391), 
SERICS (PE00000014) under the NRRP MUR program funded by the European Union - NextGenerationEU, project CRESP from the Italian Ministry of Health under the program CCM 2022, PON project “Ricerca e Innovazione” 2014-2020 and project SEED n. SP122184858BEDB3.

We would like to thank Maurizio Tesconi for providing the dataset employed for the case study.

\bibliographystyle{unsrt}
\bibliography{bibliography}

\end{document}